\documentclass[12pt]{article}
\usepackage[colorlinks=true,pdfpagemode=none,urlcolor=blue, linkcolor=blue,citecolor=blue]{hyperref}
\usepackage{amssymb,amsmath,amsfonts,amssymb}
\usepackage{graphics,graphicx,color,verbatim}
\usepackage{color}

\def\@abssec#1{\vspace{.05in}\footnotesize \parindent .2in
{\bf #1. }\ignorespaces}

\graphicspath{{/EPSF/}{Figures/}}
\DeclareGraphicsExtensions{.eps}

\def \Rm {\mathbb R}
\def \Nm {\mathbb N}
\def \Cm {\mathbb C}

\newcommand{\eps}{\varepsilon}

\newcommand{\dsum}{\displaystyle\sum}

\newcommand{\cout}[1]{}

 \renewcommand{\arraystretch}{1.5}

\begin{document}
\title{Explicit Reconstructions in QPAT, QTAT, TE, and MRE}

\author{Guillaume Bal \thanks{Department of Applied Physics and Applied Mathematics, Columbia University, New York NY, 10027; gb2030@columbia.edu} }


 
\maketitle


\begin{abstract}

Photo-acoustic Tomography (PAT) and Thermo-acoustic Tomography (TAT) are medical imaging modalities that  combine the high contrast of radiative properties of tissues with the high resolution of ultrasound. In both modalities, a first step concerns the reconstruction of the radiation-induced source of ultrasound.  Transient Elastography (TE) and Magnetic Resonance Elastography (MRE) combine the high elastic contrast of tissues with the high resolution of ultrasound and magnetic resonance, respectively. In both modalities, a first step concerns the reconstruction of the elastic displacement.

The result of this first step, which is not considered in this paper, is the availability of internal functionals of the unknown tissue properties. 
All imaging modalities are recast as the reconstruction of parameters in elliptic equations from knowledge of solutions to such equations. This paper provides a characterization of the parameters that may or may not be reconstructed from such internal functionals. We provide explicit reconstruction procedures and indicate how stable they are with respect to errors in the available measurements.

The modalities PAT, TAT, TE, and MRE allow us to reconstruct high-contrast optical and elastic properties of tissues with the high resolution of ultrasound or magnetic resonance imaging. They provide a means to reconstruct second-order tensors modeling tissue anisotropy as well as complex-valued coefficients modeling absorbing and dissipative effects.

\end{abstract}
 

\renewcommand{\thefootnote}{\fnsymbol{footnote}}
\renewcommand{\thefootnote}{\arabic{footnote}}

\renewcommand{\arraystretch}{1.1}





\section{Introduction}
\label{sec:intro}


The optical and electrical properties of tissues often display a high contrast between healthy and non-healthy tissues \cite{Arridge99,AS-IP-10,B-IP-09,CIN-SIAMREV99}. However, the reconstruction of such coefficients from boundary measurements is mathematically an ill-posed problem. As a result, modalities such as Optical Tomography and Electrical Impedance Tomography are typically low-resolution. In recent years, several methods have been developed to combine the high contrast of optical- and electrical- based modalities with the high resolution of other, often low-contrast, modalities such as ultrasound tomography or magnetic resonance imaging. Photo-acoustic tomography (PAT) and thermo-acoustic tomography (TAT) are such modalities combining high optical contrast with high ultrasound resolution.   

The physical mechanism that allows the coupling between radiation and ultrasound in PAT and TAT is the photo-acoustic effect, which may be described as follows. In both PAT and TAT, pulsed radiation propagating through a domain of interest is partially absorbed. The thermal expansion resulting from the absorption of radiation is responsible for the emission of ultrasound. An array of transducers then records the emitted ultrasound at the boundary of the domain. This is the available information from which we aim to reconstruct the optical coefficients of the tissues.

In PAT, radiation is typically near-infra-red light, while in TAT, radiation is typically low frequency (on the order of hundreds of MHz) electromagnetic radiation \cite{KRK-MP-99,S-SP-2011,WW-W-07}. In both PAT and TAT, a first step of the reconstruction procedure consists of reconstructing the amount of emitted ultrasound from the available measurements. Mathematically, this is an inverse wave problem that aims to reconstruct the initial condition from boundary time-dependent measurements. We assume this first step done; see e.g.: \cite{CLB-SPIE-09,KLFA-MP-95,WW-W-07,W-CRC-09,XW-RSI-06} for the bio-engineering literature and \cite{KK-EJAM-08,S-SP-2011,SU-IO-12,WA-SP-11} for the mathematical literature. 

The maps reconstructed during the first step provide important qualitative information. However, they do not quantify the optical properties of the tissues. This paper is concerned with the second, quantitative, step of PAT and TAT, called QPAT and QTAT, respectively; see, e.g.,  \cite{CLB-SPIE-09,CLB-BOE-10,RN-PRE-05,Z-AO-10,J-PTRSA-09} in the bio-medical literature and \cite{BR-IP-11,BR-IP-12,BRUZ-IP-11,BU-CPAM-12,BU-IP-10,GaZhOs-Prep10} in the mathematical literature. 

From the analyses in, e.g., \cite{BR-IP-11,BU-CPAM-12}, we know that not all of the unknown tissue parameters may be reconstructed in QPAT experiments. This paper follows recent results in \cite{BU-CPAM-12} to obtain a precise, general, description of all that can be reconstructed from QPAT and QTAT data.  We also describe the reconstructions in the elasticity-based imaging modalities Transient Elastography (TE) and Magnetic Resonance Elastography (MRE) when they are modeled by a scalar equation; see \cite{GCCF-JASA-03,Kal-PMB-00,LMcL-IP-09,MZM-IP-10,MLRGE-S-95,OCPYL-UI-91,PLKK-PNAS-11,STCF-IEEE-02}.

 

The elliptic models describing radiation propagation and the internal functionals resulting from solving the first step in QPAT and QTAT are presented in section \ref{sec:QPATTAT}. 
The analysis of what may or may not be reconstructed from such information is carried out in section \ref{sec:rec}. The main mathematical properties that allows us to carry out the analysis is that the {\em ratio} of internal functionals is equal to the ratio of the elliptic solutions. This property is satisfied by QPAT, QTAT, TE, and MRE. What may be reconstructed from ratios is presented in section \ref{sec:ratios}. 
The final step of the reconstruction depends on the modality of interest. QPAT is addressed in section \ref{sec:qpat}, QTAT in section \ref{sec:qtat}, and TE and MRE in section \ref{sec:TEMRE}. In QPAT and QTAT, not all the coefficients can be uniquely reconstructed. We consider several scenarios that lead to unique reconstructions. 
We stress that {\em anisotropic} as well as {\em complex-valued} coefficients can be reconstructed for all modalities.
The results are discussed in the concluding section \ref{sec:conclu}. 

QPAT, QTAT, TE and MRE belong to a class of imaging modalities now often referred to as hybrid inverse problems or coupled-physics inverse problems. 
We refer the reader to e.g. \cite{A-Sp-08,B-IO-12,KS-arxiv-11,S-SP-2011} for recent references on the rapidly evolving field of hybrid inverse problems. 

\section{Quantitative PAT and TAT}
\label{sec:QPATTAT}

The first step of QPAT and QTAT concerns the reconstruction of the map of ultrasound generated by the propagating radiation. We assume this first step done; see \cite{CLB-SPIE-09,KK-EJAM-08,S-SP-2011,SU-IO-12,WW-W-07,W-CRC-09,XW-RSI-06} for an up-to-date list of references on this problem.  The second step of QPAT and QTAT provides quantitative reconstructions of the optical parameters from knowledge of the map obtained in the first step. The mathematical problem associated to this second step is presented in section \ref{sec:modelqpat} for QPAT and in section \ref{sec:modelqtat} for QTAT. 

\subsection{Quantitative Photo-Acoustic Tomography}
\label{sec:modelqpat}

In PAT, the propagation of radiation is modeled by the following diffusion equation
\begin{equation}
\label{eq:diffPAT}
-\nabla\cdot \gamma \nabla u_j + \sigma u_j = 0 \quad \mbox{ in } X, \qquad u_j=f_j\quad\mbox{ on } \partial X.
\end{equation}
Here $u_j$ is the spatial photon density and $(\gamma,\sigma)$ are the diffusion tensor and absorption coefficient, respectively. The boundary conditions $f_j$ for $1\leq j\leq J$ indicate the different ways the domain is probed. Here, we use Dirichlet conditions to simplify. The reconstruction procedures do not depend on the choice of boundary conditions, which can thus be replaced by Neumann or more general Robin (mixed) conditions. $X$ is a regular, open, bounded, domain in $\Rm^n$ where $n\geq1$ is spatial dimension.
As radiation propagates through the domain, the resulting acoustic signal is characterized by
\begin{equation}
\label{eq:measQPAT}
  H_j(x) = \Gamma(x) \sigma(x) u_j(x).
\end{equation}
Here, $\Gamma$ is the Gr\"uneisen coefficient, quantifying the coupling between the absorbed radiation energy $\sigma(x) u_j(x)$ and the amount of resulting acoustic signal.

The general problem of QPAT is therefore to know what can be reconstructed in 
\begin{equation}
\label{eq:unknownQPAT}
  (\gamma, \sigma, \Gamma)
\end{equation}
from knowledge of $H_j(x)$ for $1\leq j\leq J$. See, e.g., \cite{BU-IP-10,CLB-SPIE-09,WW-W-07} for additional details.

\subsection{Quantitative Thermo-Acoustic Tomography}
\label{sec:modelqtat}

In TAT,  the propagation of radiation is modeled by Maxwell's equations
\begin{equation}
\label{eq:maxwell}
\nabla\times E = - \partial_t B,\quad \nabla \times H= J+\partial_t D, 
\end{equation}
where $(E,B,H,J,D)$ are the electric field, the magnetic field, the electrical displacement field, the electrical current density, and the magnetic flux density, respectively. For linear media, which we assume here, we have the relations
$J=\sigma E$, $D=\eps E$, and $B=\mu H$,
where $\sigma(x)$ is the conductivity, $\eps(x)$ the relative permittivity, and $\mu(x)$ the relative permeability.
After eliminations, we find the equation for the electric field
\begin{equation}
\label{eq:forE}
 -\nabla\times \mu^{-1} \nabla\times E = \sigma\partial_t E +\eps\partial^2_t E.
\end{equation}
Replacing the above system of equations by a {\em scalar model}, we obtain formally the scalar equation
\begin{math}
\nabla\cdot\mu^{-1} \nabla u = \sigma\partial_t u + \eps \partial^2_t u.
\end{math}
Here $u$, may be seen as one component of the electric field in a simplified setting. The above derivation may be justified when $\nabla\cdot E=0$ and $\mu$ is constant. See \cite{BRUZ-IP-11} for partial results for \eqref{eq:forE}. 
With time-harmonic sources and solutions with frequency $\omega$, the equation for $u$ becomes the Helmholtz equation
\begin{equation}
\label{eq:Helm}
\nabla\cdot\mu^{-1} \nabla u + (\eps \omega^2 -i\omega\sigma) u =0 \quad \mbox{ in } X,\qquad u=f \quad \mbox{ on } \partial X.
\end{equation}
As for PAT, more general boundary conditions can also be considered.

As radiation propagates through $X$, the emitted acoustic signal is described by
\begin{equation}
\label{eq:measQTAT}
  H(x) = \Gamma(x) \sigma(x) |u|^2(x).
\end{equation}
As above, the Gr\"uneisen coefficient $\Gamma$ describes the coupling between the absorbed radiation energy $\sigma(x) |u|^2(x)$ and the resulting acoustic signal.

Let us now assume that the medium is probed in $K$ different ways $f_j$ for $1\leq j\leq K$. Solutions of the form $|u_j|^2$ are sufficient to generate information of the form $u_ju_k^*$. Indeed, using the polarization formula for the inner product on $\Cm$, we obtain that 
\begin{equation}
\label{eq:polarization}
u_k u_j^* = \dfrac12\Big(|u_j+u_k|^2-i|u_j+iu_k|^2 - (1-i) (|u_j|^2+|u_k|^2)\Big).
\end{equation}
Therefore, using four physical experiments with boundary conditions $f_j$, $f_k$, $f_j+f_k$, $f_j+if_k$, we can reconstruct $\Gamma(x) \sigma(x)|u_j|^2$, $\Gamma(x) \sigma(x)|u_k|^2$, as well as $\Gamma(x) \sigma(x) u_k u_j^*$.
We thus assume the availability of the internal functionals
\begin{equation}
\label{eq:HjQTAT}
H_j(x) = \Gamma(x) \sigma(x) u_j u_1^*, \qquad 1\leq j\leq J.
\end{equation}
We refer to the above information $(H_j)$ as a $J-$dimensional data set even though the number of physical experiments $K$ necessary to acquire $H_j$ may be larger than $J$. The inverse problem of QTAT therefore consists in understanding what can be reconstructed in 
\begin{equation}
\label{eq:unknownQTAT}
  (\mu^{-1} (x), \eps (x), \sigma(x), \Gamma(x)),
\end{equation}
from knowledge of $(H_j)_{1\leq j\leq J}$. See e.g. \cite{AGJN-QTAT-12,BRUZ-IP-11,WW-W-07} for additional information about this model. 

\subsection{General setting for QPAT and QTAT}
\label{sec:setting}

QPAT and QTAT thus aim to reconstruct the coefficients $(a,c,\Gamma)$ from knowledge of  
\begin{eqnarray}
\label{eq:mHQPAT}
H_j(x) &=& \Gamma(x) c(x) u_j(x) \quad\qquad \mbox{ in the QPAT setting,}\\[0mm]
H_j(x) &=& \Gamma(x) \Im c(x) u_j u_1^*(x) \qquad \mbox{ in the QTAT setting,}
\end{eqnarray}
where $u_j$ is the solution to the elliptic equation
\begin{equation}
\label{eq:ellipticgal}
\nabla\cdot a\nabla u_j  + c u_j =0, \quad \mbox{ in } X,\qquad u_j=f_j\quad\mbox{ on } \partial X,\qquad 1\leq j\leq J.
\end{equation}
Here, $\Im c$ is the imaginary part of $c$. The coefficients $(a,c)$ are possibly complex-valued.

\section{Reconstruction procedure}
\label{sec:rec}

The functionals in QPAT, QTAT, TE, and MRE share the property that 
\begin{equation}
\label{eq:ratios}\dfrac{H_j}{H_k}=\dfrac{u_j}{u_k}.
\end{equation}
We exploit this structure to obtain preliminary reconstructions in section \ref{sec:ratios}. The final step of the reconstruction is modality-specific. QPAT is treated in section \ref{sec:qpat} while QTAT is handled in section \ref{sec:qtat}. 
The elasticity-based modalities TE and MRE are described in section \ref{sec:TEMRE}.

\subsection{Reconstruction from solution ratios}
\label{sec:ratios}

We assume that experiments are carried out for $J\in\Nm$ boundary conditions $f_j$, $1\leq j\leq J$, with resulting solutions $u_j$ of \eqref{eq:ellipticgal}. In the first step of the reconstruction, we assume the availability of ratios of solutions. We make our {\em first assumption} that 
\begin{equation}
\label{eq:assumption1}
u_1 \mbox{ {does not vanish} throughout the domain $\bar X$},
\end{equation}
and define  
\begin{equation}
\label{eq:vj}
v_j = \dfrac{u_{j+1}}{u_1}, \qquad 1\leq j\leq J-1.
\end{equation}
Since the functionals $H_j$ are linear in the solutions $u_j$ in QPAT and are bilinear in $(u_1^*,u_j)$ in QTAT, we observe that $v_j$ is {\em known} in the QPAT and QTAT settings. 

Some straightforward algebra shows that 
\begin{equation}
\label{eq:equvj}
\alpha : \nabla^{\otimes 2} v_j + (\nabla\cdot\alpha)  \nabla v_j =0\quad\mbox{ in }\quad\mbox{X}, \qquad \alpha(x):=a(x)u_1^2(x).
\end{equation}
Here, we use the notation: 
\begin{math}
\alpha: \nabla^{\otimes 2} v_j = \sum_{k,l=1}^n \alpha_{kl} \partial_{x_k}\partial_{x_l} v_j = {\rm Tr} (\alpha \nabla^{\otimes 2} v_j).
\end{math}
We present the reconstruction of $\alpha$ from knowledge of a sufficient number of solutions $v_j$ following \cite{BU-CPAM-12}. 
\\[2mm]
\noindent{\bf Reconstructions in the presence of a scalar diffusion coefficient.}
Let us assume that $a$ is scalar. 
Then \eqref{eq:equvj} is equivalent to the equation
\begin{equation}
\label{eq:au12}
\Delta v_j + \dfrac{\nabla (au_1^2)}{au_1^2} \cdot\nabla v_j =0.
\end{equation}
We make our {\em second assumption}:
\begin{equation}\label{eq:assumption2}
  (\nabla v_j)_{1\leq j\leq n} \quad \mbox{ forms a basis of } \Rm^n \mbox{ at each point $x\in \bar X$}.
\end{equation}
We define $H_{ij}=\nabla v_i\cdot\nabla v_j$ a matrix that is therefore invertible and $H^{ij}$ the entries of the matrix $H^{-1}$. For any complex-valued $n$-dimensional vector $F$, we then have the decomposition
\begin{equation}\label{eq:decF}
 F = H^{ij} F\cdot\nabla v_j  \nabla v_i.
\end{equation}
Here and below, we use the convention of summation over repeated indices ($i$ and $j$ above summed between $1$ and $n$).  Therefore, with ${(e_m)}_{1\leq m\leq n}$ the standard orthonormal basis in $\Rm^n$, we have
\begin{equation}
\label{eq:phiim}
e_m = \phi^m_i \nabla v_i \qquad \mbox{ for } \qquad \phi^m_i = H^{ij} e_m\cdot\nabla v_j \nabla v_i.
\end{equation}
This implies that for all $1\leq m\leq n$, we have
\begin{displaymath}
  e_m\cdot  \dfrac{\nabla (au_1^2)}{au_1^2} + \phi^m_j \Delta v_j=0,\quad \mbox{ i.e., } \quad
  \dfrac{\nabla (au_1^2)}{au_1^2}  = - \phi^m_j \Delta v_j e_m.
\end{displaymath}
In other words, the vector  $\frac{\nabla (au_1^2)}{au_1^2}$ is {\em known} explicitly. This is a redundant system of first-order equations for $au_1^2$. For instance, choosing $1\leq m\leq n$, we can solve the ordinary differential equation
\begin{displaymath}
  e_m \cdot\nabla (au_1^2) + (\phi^m_j \Delta v_j) (au_1^2)=0,
\end{displaymath}
provided that $au_1^2$ is known on $\partial X$. This provides an explicit reconstruction procedure for $au_1^2$. 
\\[2mm]
\noindent{\bf Reconstruction of a general, possibly anisotropic, diffusion tensor.}
The reconstruction when $a$ is a full tensor requires a larger number of ratios $v_j$.  We assume again that $(\nabla v_1,\ldots \nabla v_n)$ is a basis of $\Rm^n$ at each point $x\in\bar X$ and define $H_{ij}=\nabla v_i\cdot\nabla v_j$ with $H^{ij}$ the entries of $H^{-1}$. Let
\begin{eqnarray}
\label{eq:MnIn}
M_n &=& \dfrac12 n(n+1) -1,\qquad I_n = n+1+M_n = \dfrac12 n(n+3),\\
\theta^m_j &=& \left\{ \begin{array}{cl} -H^{jk}\nabla v_{m+n} \cdot\nabla v_k \quad & 1\leq j\leq n \\ 1 & j=n+m \\ 0 & \mbox{ otherwise } 
\end{array}\right. \qquad 1\leq j\leq I_n-1,\,\, 1\leq m\leq M_n.
\end{eqnarray}
These expressions for the coefficients $\theta^m_j$ ensure that
\begin{math}
  \sum_{j=1}^{I_n-1} \theta^m_j \nabla v_j =0.
\end{math}
Let us now construct the symmetric matrices
\begin{equation}
\label{eq:Mm}
M^m = \dsum_{j=1}^{I_n-1} \theta^m_j \nabla^{\otimes 2} v_j, \qquad 1\leq m\leq M_n.
\end{equation}
We deduce from \eqref{eq:equvj}  that 
\begin{equation}
\label{eq:constalpha}
\alpha : M^m = {\rm Tr}(\alpha M^m)=0 ,\qquad 1\leq m\leq M_n,
\end{equation}
for all $x\in X$. We now make our {\em third assumption}:
\begin{equation}\label{eq:assumption3}
\mbox{ We assume that $(M^m)_{1\leq m\leq M_n}$ forms a free family of symmetric matrices}.
\end{equation}
Since the dimension of the linear space of symmetric matrices is $M_n+1=\frac12 n(n+1)$, we deduce from \eqref{eq:constalpha} that $\alpha$ is in the orthogonal complement to the span of the $M_n$ matrices $M^m$, which is a one-dimensional space. Let us call $M^0(x)$ a non-trivial matrix in that space. Such a matrix $M^0$ may be obtained by Gram-Schmidt orthogonalization \cite{golub-vanloan} for instance observing that $(I,(M^m)_{1\leq m\leq M_n})$ forms a basis of the linear space of complex-valued, symmetric, matrices. This proves that 
\begin{equation}
\label{eq:alphaM0}
\alpha(x) = au_1^2(x) = \tau(x) M^0(x),
\end{equation}
for a scalar function $\tau(x)$ to be determined. Using \eqref{eq:decF} and \eqref{eq:equvj}, we deduce that 
\begin{displaymath}
\nabla\cdot (\tau M^0) = -\tau H^{ij} M^0:\nabla^{\otimes 2}v_j \nabla v_i. 
\end{displaymath}
Since $a$, and hence $M^0$, is invertible, we can recast the above equation as
\begin{equation}
\label{eq:tau}
  \nabla \tau + (M^0)^{-1} \big(\nabla\cdot M^0 + H^{ij} M^0:\nabla^{\otimes 2}v_j \nabla v_i\big) \tau=0. 
\end{equation}
This is a redundant system of linear first-order equations for $\tau$ as in the case of $\alpha$ scalar. Knowledge of $\tau$ at one point, for instance on $\partial X$, allows one to uniquely and stably reconstruct $\tau$ on $\bar X$. We have used $I_n-1=n+M_n$ ratios $(v_j)_{1\leq j\leq I_n-1}$ to obtain a unique, explicit, and stable reconstruction of the tensor $a u_1^2$.
\\[2mm]
\noindent{\bf Additional information does not provide new independent information.}
At this stage, we have reconstructed $\alpha=au_1^2$. Note that $au_j^2=au_1^2v_{j-1}^2$ is known as well for $2\leq j\leq J$.

We now show that additional internal functionals do not provide any new information when $H_ju_k=H_ku_j$. 
Indeed, let $u_k$ correspond to a new boundary condition $f_K$ and let $H_k$ be the corresponding internal functional. Then we find that
\begin{equation}
\label{eq:meask}
\nabla\cdot au_1^2  \nabla \dfrac{H_k}{H_1} =0 \quad \mbox{ in } X,\qquad \dfrac{H_k}{H_1}  = \dfrac{f_k}{f_1} \quad \mbox{ on } \partial X.
\end{equation}
This is an elliptic equation on $X$ with known Dirichlet conditions on $\partial X$. As a consequence, since $\alpha=au_1^2$ is known, then $H_k$ can uniquely be determined from the above equation. {\em There is no need to acquire $H_k$ experimentally}.
\\[2mm]
\noindent{\bf A change of variables.}
All the QPAT, QTAT, TE or MRE information is thus encoded in $(au_1^2,H_1)$. We wish to recast $au_1^2$ as a more explicit functional of the unknown coefficients $(a,c)$. Let decompose $a$ as $a=B^2\hat a$ with $\hat a$ a, possibly complex-valued, diffusion tensor such that ${\rm det}\, \hat a=1$. The amplitude of $a$ is written as $B^2$, where $B$ could again be a complex-valued scalar. It is defined uniquely by continuity on the simply connected domain $X$. 
Since $au_1^2$ is known and $u_1$ is a scalar, then $\hat a$ is known as well. 
Let us now define $v=Bu_1$. Some algebra shows that 
\begin{equation}
\label{eq:v}
\Delta v =qv \quad \mbox{ on } X, \qquad q = \dfrac{\nabla \cdot \hat a\nabla B}{B} + \dfrac{c}{B^2}.
\end{equation}
Note that $v^2=B^2u_1^2$ is known since $v^2\hat a=au_1^2$, which is known. As a consequence, $q$ is known.
\\[2mm]
\noindent{\bf A summary.}
At this stage, we have used the equations for $u_j$ and the information $v_j$ for $1\leq j\leq J-1$ to reconstruct $au_1^2$. The number of necessary internal functionals is $J=n+1$ in the case of a scalar coefficient $a$ and $J=I_n=\frac12 n(n+3)$ in the case of a symmetric tensor $a$.

All internal functionals $H_k$ can be reconstructed from knowledge of $H_1$ and $au_1^2$, and therefore no additional information can be obtained by acquiring more measurements. We have then decomposed $a=B^2\hat a$ and used the equation for $u_1$ to eliminate it and derive knowledge of $q$ in \eqref{eq:v}.  All the QPAT, QTAT, TE, or MRE information available about the coefficients is encoded in $(\hat a,q,H_1)$. What we can extract from $(\hat a,q,H_1)$ now depends on the modality under consideration.

\subsection{Reconstructions in QPAT}
\label{sec:qpat}

We recall that 
\begin{math}
 H_1(x) = \Gamma(x)  c(x) u_1(x).
\end{math}
Knowledge of $(\hat a,q,H_1)$ is thus equivalent to that of
\begin{equation}
\label{eq:coefsQPAT}
\big(\hat a,\chi,q\big)  = \Big(\hat a, \,\dfrac{\Gamma c}{B}, \,\dfrac{\nabla \cdot \hat a\nabla B}{B} + \dfrac{c}{B^2}\Big).
\end{equation}
The above information $(\hat a,\chi,q)$ (i) has been reconstructed uniquely and stably from the available data; and (ii) is all that can be obtained about $(a,c,\Gamma)$. This is the main result of this paper concerning QPAT. Note that two scalar functions, $(\chi,q)$ are known while three scalar functions $(B,c,\Gamma)$ are unknown. It is therefore impossible to reconstruct all of $(a,c,\Gamma)$ from QPAT data without further prior assumptions.  


Following the derivation in \cite{BR-IP-11}, we obtain that knowledge of one function in $(B,c,\Gamma)$ uniquely and stably determines the other two functions. For instance, if the Gr\"uneisen coefficient is assumed to be known, then \eqref{eq:coefsQPAT} provides the following elliptic equation for $B$:
\begin{equation}
\label{eq:B}
\nabla \cdot \hat a \nabla B -q B + \dfrac{\chi}\Gamma =0 \quad \mbox{ in } X,
\end{equation}
with known boundary conditions on $\partial X$. As in\cite{BR-IP-11}, we can show that the above elliptic equation admits a unique solution.  Alternatively, as also shown in \cite{BR-IP-11}, we find the equation
\begin{equation}
\label{eq:um1}
\nabla\cdot au_1^2 \nabla \dfrac{1}{u_1} + \dfrac{\chi}\Gamma =0 \quad \mbox{ in } X, \quad \dfrac{1}{u_1}=\dfrac1{f_1}\mbox{ on } \partial X.
\end{equation}
Since $au_1^2$ is known, this is an elliptic equation for $u_1$ from which we then easily deduce $a$ and $c$.

We thus obtain that knowledge of $\Gamma$ and QPAT data uniquely and explicitly determines the coefficients $(a,c)$. Note that $a$ is a possibly anisotropic tensor. When $\Gamma$ is not known, then the anisotropy $\hat a$ can still be explicitly reconstructed. However $(B,c,\Gamma)$ are reconstructed up to any transformation that leaves $(\chi,q)$ above invariant; see however \cite{BR-IP-12}.

\subsection{Reconstructions in QTAT}
\label{sec:qtat}

In QTAT, the coefficient $c$ is naturally complex valued. In the general case, $a$ is also possibly complex-valued while $\Gamma$ is a positive, real-valued, coefficient. From 
\begin{math}
 H_1(x) = \Gamma(x) \Im c( x) |u_1|^2(x)
\end{math}
we obtain that the available information about $(a,c,\Gamma)$ is 
\begin{equation}
\label{eq:coefsQTAT}
\big(\hat a,\chi,q\big) =   \Big( \hat a, \,\Gamma\dfrac{\Im c }{|B|^2}, \, \dfrac{\nabla \cdot \hat a\nabla B}{B}+ \dfrac{c}{B^2} \Big),
\end{equation}
and this is all that can be reconstructed from QTAT data.
Thus, the coefficients $(a,c,\Gamma)$ can be reconstructed up to any transformation that leaves the above coefficients $(\hat a,\chi,q)$ invariant.


When $a$ is {\em real-valued}, then the imaginary part of $q$ is $\Im c B^{-2}$. This means that {\em the Gr\"uneisen coefficient $\Gamma$ is then uniquely determined}. 
The other coefficients $(B,\Re c,\Im c)$ are determined only up to any transformation that leaves $q$ invariant. 

In the QTAT setting, $a=\mu^{-1}$. In practice, $\mu$ is always assumed to be a constant, known, scalar, parameter. In that setting, the available information is 
\begin{equation}
\label{eq:QTATmu}
   (\Gamma \Im c, \Re c,\Im c) = (\Gamma \omega\sigma, \omega^2\eps,\omega\sigma).
\end{equation}
We thus obtain that the three coefficients $(\Gamma,\eps,\sigma)$ can uniquely be reconstructed from $J=n+1$ QTAT functionals when $\mu$ is known. 

The stable reconstruction of $\sigma$ from $J=1$ functional provided that $(\Gamma,\eps)$ are known was proved in \cite{BRUZ-IP-11}; see also \cite{AGJN-QTAT-12} for an explicit, less stable, formula from $J=n+1$ functionals. 

\subsection{Reconstructions in TE and MRE}
\label{sec:TEMRE}

The mathematical tools presented above may be used to reconstruct anisotropic, complex-valued, coefficients in the imaging modalities Transient Elastography (TE) and Magnetic Resonance Tomography (MRE).  As induced elastic waves propagate through the domain $X$, the resulting displacements are imaged either by ultra-fast ultrasound tomography in TE or by magnetic resonance imaging in MRE \cite{GCCF-JASA-03,Kal-PMB-00,LMcL-IP-09,MZM-IP-10,MLRGE-S-95,OCPYL-UI-91,PLKK-PNAS-11,STCF-IEEE-02}. Assuming a scalar model for the elastic displacement, $u$ is a solution to the elliptic model \eqref{eq:ellipticgal} and 
we can consider that 
\begin{math}
H_j(x)=u_j(x).
\end{math}

As a consequence, the reconstruction of $au_1^2$ yields that of $a$. Once $a$ and $u_1$ are known, it is then straightforward to use \eqref{eq:ellipticgal} and get
\begin{equation}
\label{eq:cTE}
c(x) = \dfrac{-\nabla\cdot a\nabla u_1}{u_1}.
\end{equation}
In the TE and MRE settings, we thus observe that general complex-valued coefficients modeling possible dispersive effects $(a,c)$ can be uniquely, explicitly, and stably reconstructed. Moreover, the coefficient $a$ is allowed to be an arbitrary (elliptic) symmetric tensor.

%
\section{Discussion}
\label{sec:conclu}
%

We have obtained in \eqref{eq:coefsQPAT} and \eqref{eq:coefsQTAT} a precise characterization of what can and cannot be reconstructed from a sufficiently large number of QPAT and QTAT experiments when radiation propagation is modeled by the scalar elliptic second-order equation \eqref{eq:ellipticgal}.
The main ingredients of the derivation are the structural property $\frac{H_j}{H_k}=\frac{u_j}{u_k}$ of QPAT and QTAT functionals and the reconstruction procedure developed in \cite{BU-CPAM-12}. Such a structure is also valid in TE and MRE, in which all the coefficients (within the approximation of a scalar model) are uniquely and explicitly reconstructed.
\\[2mm]
\noindent{\bf Number of measurements.}
When the three {\em assumptions} \eqref{eq:assumption1}-\eqref{eq:assumption2}-\eqref{eq:assumption3} above are satisfied, $J=n+1$ functionals are necessary to perform the reconstruction when the diffusion coefficient is scalar, whereas $J=I_n=\frac12n(n+3)$ when $a$ is a tensor. 
In fact, the reconstructions generalize \cite{BU-CPAM-12} to the setting 
\begin{equation}\label{eq:ellipticmoregal}
 \nabla\cdot a \nabla u + b\cdot\nabla u + c u =0 \qquad x\in X,
\end{equation}
with $b$ a complex-valued vector field. The above equation is not modified when $(a,b,c)$ is replaced by $(\tau a,\tau b-a\nabla\tau, \tau c)$ for an arbitrary non-vanishing scalar function $\tau$. As a consequence, the number of degrees of freedom we can reconstruct in $(a,b,c)$ equals $\frac12 n(n+1) + n + 1 -1 = I_n$. The number of internal functionals $J=I_n$ is therefore natural in that setting and we obtain that $I_n$ functionals precisely allow us to reconstruct $I_n$ explicit functionals of the coefficients. 


The required number of internal functionals $I_n$ is therefore optimal when $b\not=0$. However, the number of measurements $I_n$ in the general case and $n+1$ in the case of $a$ a scalar, may not be optimal. For instance, one coefficient in QTAT is reconstructed from one internal functional in \cite{BRUZ-IP-11} and two coefficients in QPAT are reconstructed from two internal functionals in \cite{BR-IP-11,BU-IP-10} independent of spatial dimension $n$. 
\\[2mm]
\noindent{\bf Constraints on the elliptic solutions and boundary conditions.}
The explicit reconstruction procedure requires that  the three {\em assumptions} \eqref{eq:assumption1}-\eqref{eq:assumption2}-\eqref{eq:assumption3} be satisfied. 
These hypotheses have been shown to hold in \cite{BU-CPAM-12} for ``well-chosen" boundary conditions $(f_1,\ldots,f_J)$ in some specific situations.  In the generality considered in the present paper, the above properties are always satisfied {\em locally} \cite{BU-CPAM-12}. The procedure presented in section \ref{sec:rec} can then always be carried out locally on subsets of $X$. The reconstruction on the whole domain $X$ may then  require more functionals $H_j$ than the number $J$; see \cite{BU-CPAM-12} for additional details.
\\[2mm]
\noindent{\bf Stability estimates.}
Hybrid inverse problems are being analyzed because they provide high resolution reconstructions. Stability estimates describe how errors in the acquisition of the functionals $H_j$ propagate into errors in the reconstructed coefficient $(\hat a,\chi,q)$.

Stability estimates depend on the number of coefficients one wishes to reconstruct. 
For instance, in the QPAT setting with $a$ and $\Gamma$ real-valued and known and $c=\sigma$ real-valued, then a very simple procedure allows us to reconstruct $\sigma$ in a stable fashion. Indeed, let us assume that $u_1\geq c_0>0$ in the domain $X$. Then $\sigma u_1=\frac{H_1}{\Gamma}$ is known and hence $u_1$ can be solved from the well posed elliptic problem \eqref{eq:ellipticgal}. It remains to evaluate $\sigma=\frac{H_1}{u_1}$ and we find the existence of a constant $C$ such that 
\begin{math}
  \|\delta \sigma \|_{\infty} \leq C \| \delta H_1\|_{\infty},
\end{math}
where $\delta H_1$ is the error in the functional $H_1$ and $\delta \sigma$ is the error in the reconstruction of the absorption coefficient. Here, errors are measured in the uniform norm $\|f\|_\infty=\sup_{x\in X} |f(x)|$ although the result also holds for other choices.

In the QTAT setting, the reconstruction of $\sigma$ from one measurement $H_1$, provided that $\Gamma$ and $\mu$ and $\eps$ are constants, has been analyzed in \cite{BRUZ-IP-11}. Again, we find  that $\|\delta\sigma\|_Y\leq C \|\delta H\|_Y$ for $Y$ a space of sufficiently smooth functions; see \cite{BRUZ-IP-11}. The explicit iterative method in \cite{BRUZ-IP-11} is based on using a Banach fixed point. An explicit reconstruction from the $n+1$ internal functionals $H_j$ was recently presented in \cite{AGJN-QTAT-12}. However, the reconstruction seems to involve a loss of three derivatives, whereas the methodology in \cite{BRUZ-IP-11} requires no such loss. With the same internal functionals, we reconstruct in this paper the three real-valued coefficients in \eqref{eq:QTATmu} with a loss of one derivative (see below).

In the QPAT setting with {\em multiple} unknown coefficients, the situation is less favorable. When $a$ is not known, then $(\chi,q)$ need to be reconstructed first. We refer to \cite{BR-IP-11,BU-IP-10} for different stability estimates  for $(\chi,q)$ when $a$ is scalar and all coefficients are real-valued. When $\Gamma$ is known, we find following \cite[Theorem 4.1]{BU-IP-10} that
\begin{equation}
\label{eq:recasPAT}
\|\delta a \|_{C^{k}(\bar X)} + \|\delta c\|_{C^k(\bar X)} \leq \|(\delta H)_j\|_{C^{k+1}(\bar X;\Rm^{2n})}
\end{equation}
for $k\geq2$. Here $\|f\|_{C^k(\bar X)}=\sup_{0\leq j\leq k, x\in \bar X}|f^{(j)}(x)|$ with $f^{(j)}$ the $j-$th derivative of $f$. The reconstruction of $(a,\sigma)$ thus involves differentiating the data $H=(H_1,H_2)$ once. Such estimates still indicate that reconstructions should be accurate. The good behavior of the reconstruction of $(a,\sigma)$ was confirmed by numerical simulations conducted in \cite{BR-IP-11,BR-IP-12}.

Stability estimates for the general reconstructions presented in section \ref{sec:rec} can be obtained following the derivation in \cite{BU-CPAM-12}. The main conclusion of such estimates is that the reconstruction a full tensor $a$ is less stable than when the coefficient $a$ is scalar. Upon inspection of \eqref{eq:Mm}, we observe that the data $v_j=\frac{H_j}{H_1}$ need to be differentiated twice when $M^m$ is constructed. The reconstruction of $e_m$ in \eqref{eq:phiim} when $a$ is scalar involves only first derivatives of the data $v_j=\frac{H_j}{H_1}$ (the estimated second derivatives are followed by one integration, which cancels the loss of one derivative). Consider the QPAT setting with $\Gamma$ known. Then we find the stability estimates:
\begin{equation}
\label{eq:estimhata}
\|\delta \hat a\|_{C^0(X)} + \|\delta \sigma \|_{C^0(X)} + \|\delta B\|_{C^1(X)} \leq C \|(\delta H)_j\|_{C^2(X;\Rm^J)}.
\end{equation}
We lose two derivatives to reconstruct $(\hat a,\sigma)$. Since $B$ is reconstructed using \eqref{eq:tau}, we actually gain one derivative after integration and the stability estimates predict a better reconstruction of the scalar component $B$ of $a$ than the absorption coefficient $\sigma$.
\\[2mm]
\noindent{\bf Numerical implementation.}
The reconstructions presented above for QPAT when only $(a,\sigma)$ is unknown and $a$ scalar and for QTAT when only $\sigma=\Re c$ is unknown have been implemented numerically in \cite{BR-IP-11,BRUZ-IP-11}. In \cite{BR-IP-11}, the stability estimate \eqref{eq:recasPAT} predicts that data need to be differentiated once. Of course, the differentiation of noisy data should not be done without some processing. A low-pass filter, for instance by a convolution with a kernel with adapted width, may for instance be applied to the data prior to differentiation. The treatment of noisy data has been the object of considerable research in the inverse problems community. We refer the reader to e.g., \cite{engl,Kirsch-IP,vogel-SIAM02} for a large class of standard methodologies to address the differentiation of noisy functions. 

\section*{Acknowledgment}
This work was partially funded by a grant from the U.S. National Science Foundation.

{\footnotesize

}
\end{document}